\documentclass[letterpaper]{article} 
\usepackage{aaai25}  
\usepackage{times}  
\usepackage{helvet}  
\usepackage{courier}  
\usepackage[hyphens]{url}  
\usepackage{graphicx} 
\def\UrlFont{\rm}  
\usepackage{natbib}  
\usepackage{caption} 
\usepackage{algorithm}
\usepackage{algorithmic}

\usepackage{enumitem}
\usepackage{newfloat}
\usepackage{listings}
\DeclareCaptionStyle{ruled}{labelfont=normalfont,labelsep=colon,strut=off} 
\floatstyle{ruled}
\newfloat{listing}{tb}{lst}{}
\floatname{listing}{Listing}
\usepackage{xcolor}

\title{Bidirectional Human-AI Learning in Real-Time Disoriented Balancing}
\author{
    Sheikh Mannan, Nikhil Krishnaswamy
}
\affiliations{
    Situated Grounding and Natural Language (SIGNAL) Lab, Colorado State University\\

    Fort Collins, CO 80523 USA\\
    \{sheikh.mannan,nkrishna\}@colostate.edu
}

\usepackage{bibentry}

\begin{document}

\maketitle

\begin{abstract}
We present a real-time system that enables bidirectional human-AI learning and teaching in a balancing task that is a realistic analogue of disorientation during piloting and spaceflight. A human subject and autonomous AI model of choice guide each other in maintaining balance using a visual inverted pendulum (VIP) display. We show how AI assistance changes human performance and vice versa.
\end{abstract}

%

\section{Introduction}

Spatial disorientation remains a leading causes of fatal aircraft accidents \cite{braithwaite1998spatial,gibb2011spatial}. An automated system, without the physical characteristics that lead to the inducement of disorientation, can potentially serve as a countermeasure \cite{wang2022crash}. However, the automated system may execute strategies counter to the human's own preference, and too many such instance may result in the human losing trust in the AI \cite{nikolaidis2017human} even when AI assistance makes overall human task performance better. One way to mitigate such conflict is {\it dyadic interaction}, wherein the process of interacting with a partner changes one's own underlying behavior \cite{park2012need,roy2017formation}.

\begin{figure}
    \centering
    \includegraphics[width=.9\columnwidth]{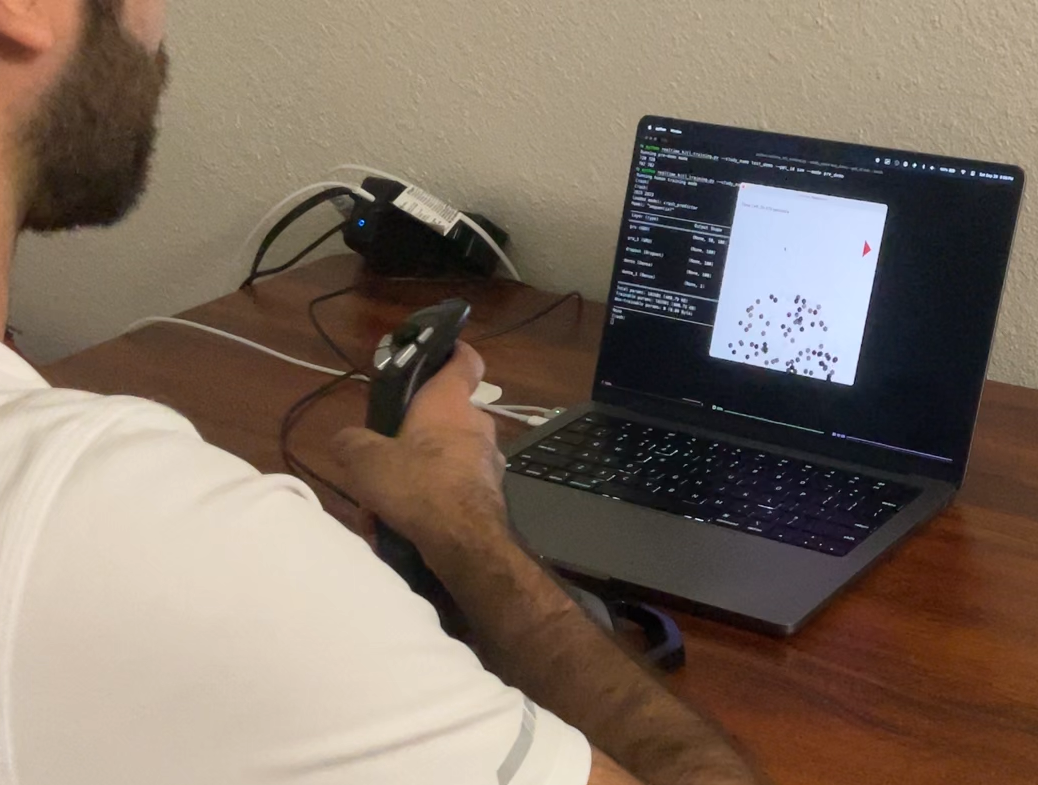}
    \caption{A human engaging in VIP balancing with cues (arrows) rendered by an AI assistant. A demonstration video can be viewed at \url{https://youtu.be/coJdj0LIYa4}.}
    \label{fig:video-recording}
\end{figure}

In this system, we demonstrate how mutual human-AI adaptation may be conducted in real-time using a realistic simulation of balancing under disorientation, where human participants use a joystick to stabilize themselves about the direction of balance (DOB). See Fig.~\ref{fig:video-recording}. A human user may be paired with different types of AI models which themselves display a variety of proficiency and performance strategies in this action-learning task with well-controlled parameters that is nonetheless challenging for humans. Human and AI mutually correct each other's actions with either visualized suggestions (AI-to-human) or direct numerical joystick input (human-to-AI). The differences in both human and AI performance before and after dyadic learning can be clearly visualized in intuitive phase portraits.

\section{Task Setting}

The \textbf{virtual inverted pendulum (VIP)} task is a documented, realistic simulation of upright balance control in a disorienting condition analogous to spaceflight or piloting in the absence of strong gravitational cues. The IP is represented as a visually simulated circular array of dots (random dot kinematogram, RDK) which rolls in the plane of the display screen governed by $\ddot{\theta} = k_{P}sin\theta$, where $\theta$ is degrees deviation from the DOB and pendulum constant $k_{P} =$ 600$^{\circ}$/s$^2$ \cite{vimal2017learning}. If the IP drifts beyond $\pm60^\circ$ from the DOB, it has ``crashed''. To induce disorientation, the VIP is rendered at 50\% {\it coherence}: between every two consecutive frames, half the dots displace coherently while the other half jump randomly. This eliminates configural displacement cues relative to the DOB while providing low-level retinal motion cues. The result is that humans watching the VIP in motion can tell how fast they are moving, but have difficulty telling how far from the upright they have fallen, as if they had been denied gravitational cues. Thus, in this high-throughput, portable setting, humans experience performance degradations in maintaining balance that strongly correlate with those induced by placement in a physical apparatus that disrupts signals from the vestibular system \cite{panic2015direction,panic2017gravitational,vimal2016learning,vimal2018learning,vimal2017learning,vimal2019learning,vimal2021role,vimal2020characterizing,wang2022crash,mannan2022and,dizio2023manual,mannan2024combating,mannan2024embodying}.


A perfect performer in this task would rotate to the DOB and remains there with no drift or oscillation. Thus a proficient performer minimizes, e.g., distance from the DOB, angular velocity, and magnitude of actions. This means we can demonstrate bidirectional learning in real-time as humans or AIs display characteristic behaviors that may be more idealized (as above) and/or more human-like (such as small intermittent deflections, as in \citet{vimal2020characterizing}) and these characteristics may converge as the two learn from each other.

\section{System Functionality}

An AI ``assistant'' can be instantiated as any of a variety of reinforcement learning (RL) or supervised deep learning models. Examples include SAC or DDPG instances trained in an environment programmed with the IP physics, or MLP, RNN, LSTM, or GRU models trained on data from humans performing disoriented IP balancing tasks using the VIP or an analogous physical apparatus \cite{panic2015direction,vimal2017learning,mannan2024combating,mannan2024embodying}. There are 26 available assistants, detailed in \citet{mannan2024combating}. All AI models predict what the next joystick deflection would be given the current angular position and velocity, and so depending on the model type and training data, the AI model may display different levels of native proficiency at performing the task.

In our system, after a tutorial to help the user acclimate to the joystick, controls, pendulum, and RDK movement, bidirectional human-AI learning proceeds in two phases:

\begin{enumerate}
    \item Human training: First, the user performs the task alone to determine baseline performance. Then, the human is assisted by an AI, which provides suggestions when deemed necessary, rendered as arrows on screen.
    \item AI training:
    \begin{enumerate}[label=\alph*.]
        \item An AI performs the task alone. During solo performance, the AI receives only numerical input, but depending on model and training data may display any level of proficiency at the task.
        \item The human then assists the AI in a second run by deflecting the joystick to keep the VIP balanced. This is done in the 50\% coherent VIP condition to ensure that signals the AI receives are those from a human who is experiencing disorientation. All episodes where the human and AI disagreed on the direction of the movement are recorded. After the run, a brief finetuning is performed to update the assistant.
        \item The updated AI performs the task again where the human can determine whether the AI has improved or requires further corrections and updates. If further corrections are required step (b) can be repeated until the AI achieves acceptable performance.
    \end{enumerate}
\end{enumerate}

\begin{figure}
    \centering
    \includegraphics[width=.495\columnwidth,trim={25 25 25 25},clip]{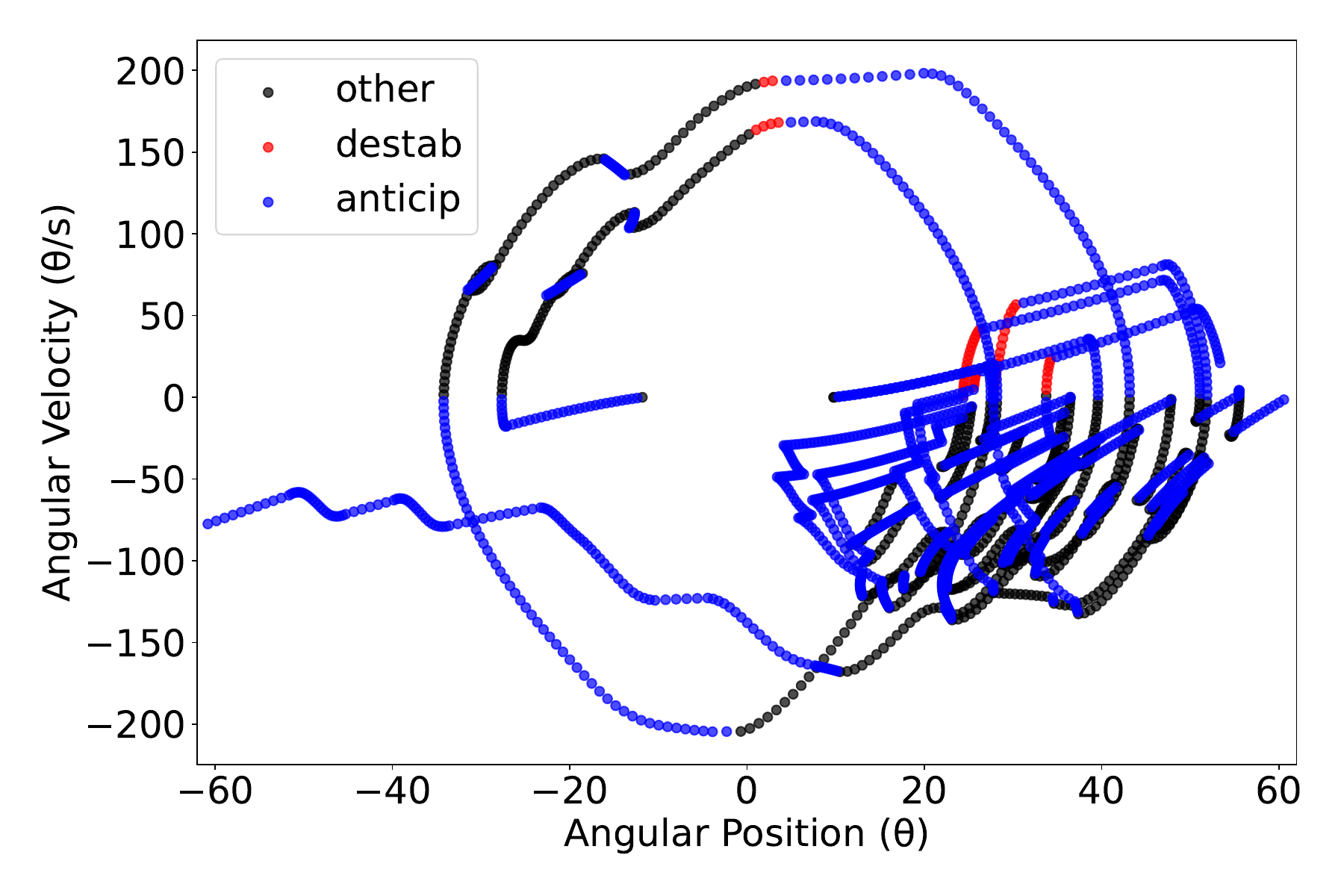}
    \includegraphics[width=.495\columnwidth,trim={25 25 25 25},clip]{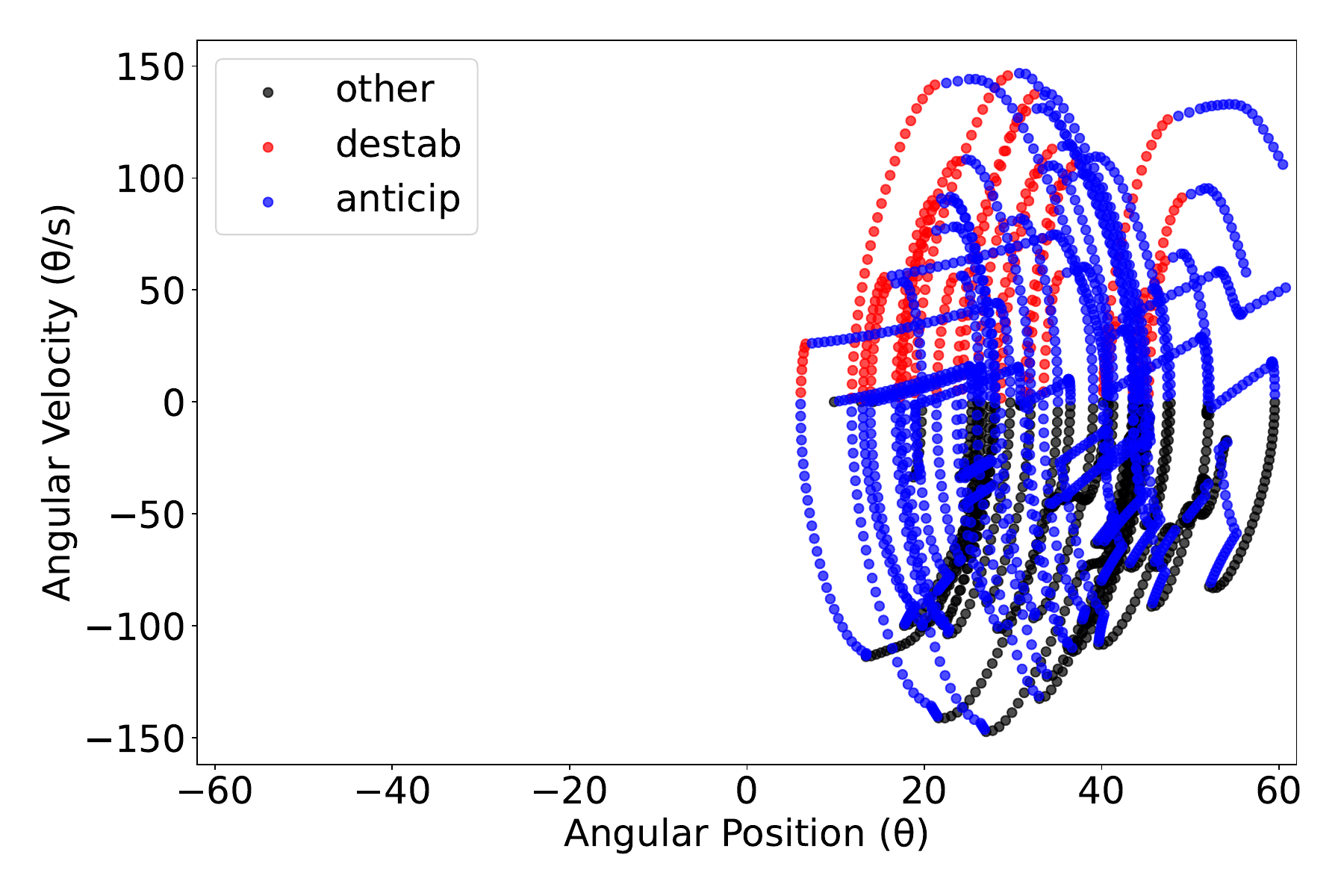}
    \caption{Phase portraits of sample human VIP performance without [L] and with [R] AI assistance. With AI assistance, this human subject decreased their oscillation and maintained stability even while offset from the DOB.}
    \label{fig:phase-plot}
\end{figure}

After each phase, the baseline and assisted performances of the human or AI are shown in the form of phase portraits of angular velocity vs. angular position (e.g., Fig.~\ref{fig:phase-plot}).

\section{Technical Specifics}

RL assistant models learned directly from exposure to environmental physics using a custom variation of Gymnasium's classic-control Pendulum environment, modified to reflect the dynamics of the VIP task. The deep learning models have been pretrained on human subject data gathered from trials performed in a physical Multi-Axis Rotation System (MARS) apparatus, programmed with identical dynamics, where subjects likewise use a joystick to balance the device while deprived of orientational cues (details in \citet{wang2022crash}). Additional data was gathered from subject trials in the VIP setting (details in \citet{mannan2024combating}).

We also incorporate a {\it crash predictor}, which is a stacked GRU model as reported in \citet{wang2022crash}, which predicts the likelihood of a crash occurring. AI cueing is provided in cases of imminent danger (crash is $\geq$80\% likely) where angular distance from the DOB exceeds 12$^\circ$.

For fine-tuning models during the AI training phase, the actor networks of the SAC and DDPG are fine-tuned using behavior cloning over the trial data, the SAC-AIRL model is updated using AIRL over the trial data, and the deep learning models undergo standard fine-tuning. The RL models are fine-tuned for 100 epochs with a learning rate of $1e-5$ and a batch size of 64. The deep learning models are fine-tuned for 20 epochs with a learning rate of $1e-7$, a 9:1 train-test ratio, and a batch size of 16. All model fine-tuning can be conducted on a consumer laptop and takes approximately 30 seconds for a single training run.

\section{Conclusion}

Our system concisely demonstrates how human-AI mutual adaptation manifests through dyadic bidirectional learning, evident in changes in both human and AI behavior due to the interaction. Our demonstration is lightweight and showcases mutual human-AI learning in near-real-time on equipment as common as a consumer laptop, making it an accessible public-facing demonstration of AI. Our task setting is directly applicable to problems of spatial disorientation as in piloting or spaceflight, but our demonstration of human-AI bidirectional learning has broader applicability to the study of shared autonomy and problems in human-AI trust and allows for rapid parameterization of multiple experiments to test hypotheses in this area. Our code is available at \url{https://github.com/csu-signal/HITL-VIP/releases/tag/v1.0}.

\section{Acknowledgments}
Thanks to Paul DiZio, Vivekanand Pandey Vimal, and James R. Lackner for foundational work on the spatial disorientation problem and collecting and providing the initial training data, to Hannah N. Davies for additional data collection, and to Paige Hansen for help training the candidate assistant models.

\bibliography{aaai25}

\end{document}